\documentclass[preprint,showpacs,preprintnumbers,amsmath,amssymb]{revtex4}

\usepackage{amsmath}
\usepackage{amsfonts}
\usepackage{mathrsfs}
\usepackage{graphicx}
\usepackage{dcolumn}
\usepackage{bm}
\usepackage[colorlinks, dvipdfm, citecolor=blue]{hyperref}
\usepackage{booktabs}
\begin{document}

\title{Constructing  silicon nanotubes by assembling hydrogenated silicon clusters}

 \author{Lingju Guo}
 \author{Xiaohong Zheng}
 \author{Chunsheng Liu}
 \author{Wanghuai Zhou}
 \author{Zhi Zeng\footnote{Corresponding author. E-mail: zzeng@theory.issp.ac.cn}}%
\affiliation{%
Key Laboratory of Materials Physics, Institute of Solid State
Physics, Chinese Academy of Sciences, Hefei 230031, P.R. China
}%

\begin{abstract}

The search or design of silicon nanostructures similar to their
carbon analogues has attracted great interest recently. In this
work, density functional calculations are performed to
systematically study a series of finite and infinite hydrogenated
cluster-assembled silicon nanotubes (SiNTs). It is found that stable
one-dimensional SiNTs with formula $Si_{m(3k+1)}H_{2m(k+1)}$ can be
constructed by proper assembly of hydrogenated fullerene-like
silicon clusters $Si_{4m}H_{4m}$. The stability is first
demonstrated by the large cohesive energies and HOMO-LUMO gaps.
Among all such silicon nanotubes, the ones built from
$Si_{20}H_{20}$($m=5$) and $Si_{24}H_{24}$($m=6$) are the most
stable due to the silicon bond angles that are most close to the
bulk $sp^{3}$ type in these structures. Thermostability analysis
further verifies that such tubes may well exist at room temperature.
Finally, both finite nanotubes and infinite nanotubes show a large
energy gap. A direct-indirect-direct band gap transition has been
revealed with the increase of the tube radius. The existence of
direct band gap may make them potential building blocks for
electronic and optoelectronic devices.


\end{abstract}

\maketitle

\section{INTRODUCTION}
Since the discovery and application of carbon fullerenes and carbon
nanotubes(CNTs) \cite{fullerene}, stable cage and tube-like
structures have attracted a great deal of attention. Silicon and
carbon are members of the same group in the periodic table,
suggesting a potential probability to form similar structures.
Furthermore, due to the fundamental importance of silicon in
present-day integrated circuits, substantial efforts have focused on
investigating nano-scale forms of silicon, both for the purpose of
further miniaturizing the current microelectronic devices and in the
hopes of unveiling new properties that often arise at the nano-scale
level\cite{JPCM_16_1373}. However, it is difficult to form cages or
tubes like carbon fullerenes or nanotubes purely with Si atoms
because silicon does not favor the $sp$$^{2}$ hybridization that
carbon does. Carbon normally forms strong $\pi$ bonds through
$sp^{2}$ hybridization, which can facilitate the formation of
two-dimensional spherical cages (or planar structures such as
benzene and graphene). Silicon, on the other hand, usually forms
covalent $\sigma$ bonds through $sp^{3}$ hybridization, which favors
a three dimensional diamond-like structure.

Interestingly, it has been reported that Si cage clusters can be
synthesized by encapsulating suitable foreign atoms to terminate the
dangling Si bonds that inherently arise in cage-like networks. Many
researchers\cite{jctn257,cms1,prb77195417} reported that transition
metal(TM) atoms are the most suitable elements for cage formation
due to their $d$ band features. In addition, rare earth atoms have
also been doped into silicon
cages\cite{jcp084711,epjd343,prb125411,prb115429}. Another way to
stabilize the Si cages is to terminate the cluster surface by
hydrogen\cite{prb075402,prb155425,prb80195417,handbook}, which is
similar to the dodecahedral C$_{20}$H$_{20}$.

Meanwhile, tube-like silicon nanostructures have also attracted
great attention. Thus far, a few hollow and nonhollow silicon
nanotube structures have been proposed based on intuition or the
behavior of similar materials and theoretically characterized in
recent years \cite{pnas2664,handbook,prb80195417,jmst127, prl265502,
nl301, nl1243, nano109, jmc555, njp78, prl146802, cpl81, prb075420,
prb195426, jpcb7577, jpcb8605, prb9994, jpcc5598, prb11593,
prb205315, prb075328, ss257, prb193409, jpcc16840, prl1958, pssr7,
prb155432, prl792, jpcc1234}. Nevertheless, most of these structures
have the instability problem arising from the unsaturated dangling
bonds. Among all the present schemes for constructing silicon
nanotubes, one is of great interest in which metal atoms are
encapsulated in the tubes. This scheme has two advantages.  On one
hand, the metal atom is able to support the tube wall so as not to
collapse. On the other hand, it can saturate the dangling bonds and
further stabilize the structure. However, this scheme is limited to
tubes with very small radii($R$$\leq$1nm) since for larger tubes,
the tube becomes metallic, then the above two advantages will
disappear.

In comparison,  hydrogen saturation outside the tubes should also be
a very good way for the construction of silicon tubes for three
reasons. At first, each dangling bond can be saturated by one
hydrogen atom. Thus the instability caused by the dangling bonds
would be removed and this scheme is not limited to small size tubes.
 Moreover, compared with the metal encapsulation scheme,
 the intrinsic features of silicon dominate in the properties of the
tubes  with no interference of metal atoms. More importantly, in
geometry, it is a ``real'' empty tube, instead of a filled one.  In
fact, the effect of outside saturation of dangling bonds has been
already demonstrated in experiments, where the surfaces of silicon
nanowires(SiNWs) are always passivated by hydrogen atoms
\cite{science1874,jpcb8605} or by silicon oxide layers
\cite{am1219,am1172,am564,prl116102}.

Therefore, hydrogen terminated silicon cages will be  perfect
building blocks for Si nanotubes and in this work, we present our
design of hydrogenated cluster assembled single wall silicon
nanotubes and systematically investigate their stabilities and
electronic properties using density functional theory(DFT)
calculations. We find that  stable one-dimensional silicon nanotubes
(SiNTs) with formula Si$_{m(3k+1)}$H$_{2m(k+1)}$ can be constructed
by proper assembly of hydrogenated fullerene-like silicon clusters
Si$_{4m}$H$_{4m}$ and these tubes can even exist at room
temperatures.

\section{COMPUTATIONAL DETAILS AND MODEL DESIGN}\label{cd}

All theoretical computations are performed with the DFT approach
implemented in the Dmol$^{3}$ package \cite{jcp92,jcp113}, using all
electron treatment and the double numerical basis including the
$d$-polarization function(DNP)\cite{jcp92}. The exchange-correlation
interaction is treated within the generalized gradient
approximation(GGA) using BLYP functional. Self-consistent field
calculations are performed with a convergence criterion of
2$\times$10$^{-5}$ Hartree on total energy. The converge threholds
are set to 0.002 Hartree/{\AA} for forces and 0.005{\AA} for the
displacement.

The single Si$_{4m}$H$_{4m}$~($m=4, 5, 6, 7, 8$) cage-like clusters
are optimized first, and some of the initial structures are based on
the results reported in Refs.~\onlinecite{prb075402} \&
\onlinecite{prb155425}. Then the optimized stable single clusters
are taken as basic units (keep them as original) and stacked
together along the axis of symmetry to construct finite nanotubes,
with the adjacent two units sharing the same bottom surface. One
dimensional infinite nanotubes are also investigated by including
the smallest repeated unit cell in the supercell, with the size
chosen as 25{\AA}$\times$25{\AA}$\times$$L$$_{z}$, where
 the direction of $z$  is defined as the axial direction of the nanotubes, and $L_{z}$
 is the length of the supercell in the z direction. Meanwhile, in order to avoid interaction from the
 adjacent tubes, a sufficiently large vacuum region is introduced along the
radial directions. The Brillouin zone was sampled with a
$1\times1\times20$ irreducible Monkhorst-Pack k-point grid for
structural relaxation and band structure calculations.

Thermal stability of the hydrogenated silicon nanotubes is studied
within $ab$ $initio$ quantum molecular dynamics framework performed
by heating at 400K by using NVT(constant volume and temperature)
dynamics with a massive Nose\'{e}-Hoover thermostat. Time step is
set as 1.0 fs, total simulation time was set as 4.0 ps.

\section{RESULTS AND DISCUSSIONS}

\subsection{Structures of Si$_{4m}$H$_{4m}$ clusters and finite nanotubes }\label{f}

The fully optimized structures of Si$_{4m}$H$_{4m}$~($m$=4, 5, 6, 7,
8) clusters  are shown in Fig. \ref{fig1}. All these structures
share the following common characteristics: 1. All of them are
fullerene-like hollow structures; 2. Each Si atom has three Si
neighbors, with one H atom saturating the dangling bond outside the
cage and thus  an $sp^3$ type hybridization is satisfied;  3. All
these structures consist of $2m$ polygons and two other polygons at
the two ends, with the edge number of these two polygons as $m$.
Meanwhile, these two polygons are parallel to each other, but with a
relative angle of $\frac{\pi}{m}$ between them. Thus each vertex
atom of one polygon falls exactly on the perpendicular bisector of
one edge in the other polygon. Specifically, for
$Si_{20}H_{20}$($m$=5), the cage is composed of 12 pentagons, which
is very similar to the structure of carbon fullerene $C_{20}H_{20}$.
 In addition, structures of $Si_{16}H_{16}$($m$=4), $Si_{24}H_{24}$($m$=6)
 and $Si_{28}H_{28}$($m$=7)
have been widely discussed in very recent years
\cite{prb075402,prb155425,handbook} and the structural information
we obtained is consistent with these reports.

Taking these original clusters  as basic units, we stack
them along the central axis of the cage to form finite nanotubes.
The two adjacent cages share the same bottom polygon. We note that for
the shared polygon, there is no need for hydrogen saturation
because each Si atom already has four  Si neighbors and thus the $sp^3$
hybridization bond type is fulfilled. The molecular formula of the
finite tube can be written  as $Si_{m(3k+1)}H_{2m(k+1)}$, where the
 number of units $k$ defines the length of the nanotube, while the number of
 atoms $m$ in the bottom polygon defines the size
in the directions vertical to the axis and can be taken as a
measurement of the radius. Consequently, each group of $m$ and $k$
uniquely defines a nanotube with different radius and length
therefore such a nanotube can be denoted as NT($m, k$). After full
optimization, for one certain value $m$($m=4, 5, 6, 7$), and for $k$
ranging from 2 to 4 concerned in the present work, the tubes are
always straight and stable. Furthermore, if the number of repeated
units $k$ is fixed, the length of the tubes decreases with the
increasing $m$. The angles of H-Si-Si and Si-Si-Si inside the
repeated units are all about 109$^{\circ}$, which is very close to
the 109.5$^{\circ}$ of $sp^3$, but the Si-Si-Si angle between two
units becomes smaller and smaller with the increase of tube radius
(from 127.8$^{\circ}$ of $m$=4 to 99.0$^{\circ}$ of $m$=8).

\subsection{Electronic structures}

In order to measure the relative stability of the tubes as well as
the influence of the length and width, we have calculated the
cohesive energy ($E_{coh}$). The $E_{coh}$ are defined by the
following formula\cite{apl203112,prl157405,prb1419}:

\begin{equation}
 \ E_{coh}=\{BE[NT(m, k)]+\mu_{H}N_{H}\}/N_{Si}
 \label{a}
\end{equation}
where BE[NT(m, k)] is the binding energy of finite
nanotube($Si_{m(3k+1)}H_{2m(k+1)}$), $N_{Si}$ and $N_{H}$ are the
number of Si and H atoms, respectively, $\mu_{H}$ is the chemical
potential of H, and thus the comparison of relative stability of
different systems becomes straightforward. This effectively removes
the energy contribution of all Si-H bonds in every system.

As illustrated in Fig. \ref{fig2}(a), the cohesive energy $E_{coh}$
of the finite tubes is above 2.95eV and increases gradually as the
length $k$ increases for each fixed $m$, which indicates that the
tube becomes increasingly stable as it gets longer. On the other
hand, the stability of the tube does not depend monotonously on the
tube radius as measured by $m$. Particularly, the $E_{coh}$ curves
of $m$=5 or $m$=6 are very close and are obviously  higher than the
others, which means that, for any length $k$, the tubes with radius
$m$=5 or $m$=6 are the most stable.

In addition, Fig.~\ref{fig2}(b) shows  the variation of energy gaps
between the highest occupied molecular orbital~(HOMO) and lowest
unoccupied molecular orbital~(LUMO) of the finite tubes. When $m$
(radius of the tube) is fixed, the HOMO-LUMO gap decreases as the
length ($k$) of the nanotube increases. Comparing the gaps of the
tubes with different radius (or $m$, see Fig. \ref{fig2}(b)), we
find that the tubes  NT($4, k$), NT($5, k$) and NT($6, k$) always
have similar HOMO-LUMO gaps due to the analogous structural
parameters characterized by bond angle and length. With the further
increase of the tube radius($m$= 7, 8), the HOMO-LUMO gap decreases
rapidly.

The spatial resolved local density of states (LDOS) of HOMO and LUMO
states of cage clusters and finite nanotubes are plotted in
Fig.~\ref{fig3} for the understanding of the bonding properties. For
the convenience of descriptions, we define the atoms of two opposite
polyhedron as bottom atoms, and the other atoms lying between the
two polyhedrons as side atoms.  For small size clusters ($m=4, 5$),
both bottom atoms and side atoms have contributions on HOMOs, but
for large size clusters ($m=6, 7, 8$), HOMOs are almost localized on
the side silicon atoms. On the other hand, The LUMO states of small
size clusters ($m=4, 5, 6$) are mainly localized inside the cage
(Fig.~\ref{fig3}(a)) while for bigger size clusters $m=7, 8$, the
LUMO states are localized around the bottom atoms. For finite
nanotubes NT($m, 4$)  with small size ($m =4, 5$), both the HOMO and
LUMO are distributed on the side atoms and  the shared bottom atoms
(see (Fig.~\ref{fig3}(b)). However, for finite tubes with large
sizes ($m = 6, 7, 8$), the HOMO is localized on the side atoms while
the LUMO is localized on the shared bottom atoms, which are very
similar to their building units, namely, the cage clusters.
Furthermore, with the increase of the radius, both the HOMO and LUMO
tend to be localized on the middle atoms while the contributions
from the end atoms become negligible.

\subsection{Infinite nanotubes}
The increased stability of finite nanotubes with increasing length
leads to our interest in examining further the stability of infinite
nanotubes. The smallest repeated unit cell of the infinite nanotubes
can be obtained  by removing the Si and H atoms of one bottom
polygon and the H atoms of the other bottom polygon of a finite tube
NT$(m,2)$ or $Si_{7m}H_{6m}$, and thus it can be described by a
formula $Si_{6m}H_{4m}$, where $m$ is again the number of atoms in
the bottom or connecting polygon. Note that the repeated unit cell
should not be based on NT$(m, 1)$ but on NT$(m, 2)$ to produce a
periodic system. Therefore we can define these infinite tubes as
NT(m, $\infty$) ($m=4,5,6,7,8$). Two repeated cells of the infinite
tubes are shown in Fig.~\ref{fig4} with different radii concerned in
this work.

Full structure relaxation indicates that the infinite nanotubes have
similar geometric structures to finite ones, but the length of the
 smallest repeated cell is slightly changed. The lengths of the
smallest repeated cell of the $m=4$ tube  and the $m=5$  tube become
0.08{\AA} and 0.11{\AA} longer than those of the finite ones for
producing these repeated cells. But for $m=6$, $m=7$ and $m=8$,  the
lengths become 0.075{\AA}, 0.212{\AA} and 0.365{\AA} shorter. The
diameter of tubes and Si-H bond lengths are almost the same as those
of the finite ones. The average Si-Si bond length is 2.401\AA, and
the Si-H bond length is 1.497\AA, which are similar to the 2.36\AA~
and 1.50\AA~in Ref.~\onlinecite{pssr7}, and the exohydrogenated
carbon nanotube like structures (2.34\AA~and 1.51\AA~in
Ref.~\onlinecite{prb193409}, 2.335\AA~and 1.521\AA~in Ref.
\onlinecite{ss257}).

In order to study the stability of the infinite tubesNT(m,
 $\infty$), their cohesive energies are calculated and included in
Fig.~\ref{fig2}(a). By comparison, we find that the cohesive
energies of them are larger than those of finite ones, which means
that it is possible to synthesize long tubes. Further, thermal
stability of the hydrogenated silicon
 tubes has been checked within $ab$ $initio$ quantum molecular dynamics
 framework performed by heating at 400K for 4.0 ps
 with the time step of 1.0 fs using NVT(constant volume and
 temperature) dynamics with a massive Nose\'{e}-Hoover
thermostat. The cluster model is implemented for these tubes without
any symmetry constraints so that  all atoms are allowed to move
freely. No collapse is found at this time scale, indicating that
these hydrogenated silicon nanotubes may survive at room
temperature.

Another concern about the structural stability of the proposed
hydrogenated silicon tubes comes with whether multiple tubes will be
collapsed and merged into a large cluster when they are put
together. Our calculations on NT(5, $\infty$) and NT(7, $\infty$)
show that when two, three or four nanotubes are placed together in
parallel, with very small initial distance between them, after full
optimization, all of these tubes separate away from each other with
no collapse or distortion. The separation is largely due to the
mutual repulsion of the surface hydrogen atoms. It gives another
proof that single hollow silicon nanotubes can well exist and they
will not be combined together with the ``protection'' from the
surface hydrogen atoms between the tubes.

Finally, the  band gaps($\Delta_{g}$) of the infinite tubes are
analyzed and exhibit large values from 2.3eV of NT(8, $\infty$) to
2.8eV of NT(4, $\infty$), which implies that they are wide gap
semiconductors. Particularly, the band gaps of NT(5, $\infty$) and
NT(6, $\infty$) are obtained as 2.69 eV and 2.71 eV, respectively,
which agree very well with the 2.65 eV and 2.70 eV reported in
Ref.~\onlinecite{pssr7} for the same structures. The band gap of the
tube is inversely proportional to the radius, which is similar to
the gap changes in exohydrogenated single-wall carbon
nanotubes(SWCNT)\cite{prb075404} and exohydrogenated single-wall
silicon nanotubes\cite{prb193409}. Meanwhile, seen from
Fig.~\ref{fig5}, the type of the band gaps can be controlled by
tuning the tube radius. The smallest tube NT(4, $\infty$)
(Fig.~\ref{fig5}(a)) has a direct band gap at Z-point, while
NT(5,$\infty$)(Fig.~\ref{fig5}(b)) has an indirect band gap. When
$m$ in NT($m$, $\infty$) goes to $m=6,7,8$, these tubes all display
a direct band gap at $\Gamma$-point. From Fig.~\ref{fig5}, we see
that for $m=4$, both the bottom of the conduction band (BC) and the
top of the valence band (VT) lie at Z point, thus the $m=4$ tube is
a semiconductor with a direct gap at Z point. Interestingly, with
the increase of the radius, both the BC and TV move toward the
$\Gamma$ point. However, the BC moves much faster than TV, thus an
indirect gap is observed with $m=5$.  After $m=6$, both BC and TV
arrive at $\Gamma$ point and a direct gap is always observed . We
believe such an band evolution is related to the structure changes
with the radius, and the nanotubes have a one-dimension-like to
three-dimension-like transition as the increasing radius. For the
thinnest nanotube NT(4, $\infty$), the confinement along Z direction
is smaller than that along vertical directions because of the one
dimension feature. Consequently the band gap at $\Gamma$ is larger
than that at Z, then the effective gap opens at Z point. On the
other hand, for the largest radius tube NT(8, $\infty$), three
dimensional feature is apparently shown and the confinement along
vertical directions is relatively smaller than that of small tubes.
This makes the band gap at $\Gamma$ be small, so that the effective
gap opens at $\Gamma$ point. Accordingly, as the radius of tube
changes from NT(4, $\infty$) to NT(8, $\infty$), the confinement
along vertical directions changes evidently. This variation leads to
a transition between direct band gap and indirect band gap. Larger
tubes NT (9, $\infty$) and NT (10, $\infty$ ) are tested, and the
band structures of the tubes are shown in supplementary materials.
We can find that they are also both semiconductors with a direct
band gap at $\Gamma$ point. Our results also show that band gap open
at $\Gamma$ is more sensitive to the change of diameter than that at
Z point. The reason for this is that the size of basic repeated unit
cell keeps almost unchanged in building nanotube, hence the physical
phenomena close to the size of basic repeated unit cell are not
sensitive to the size change of nanotube. In fact, such a size
induced change is also observed in the HOMO-LUMO gap and the
cohesive energy in the finite tubes (see Fig.~\ref{fig2}). In
Fig.~\ref{fig2}, we see that $m=5$ is a special size, since after
$m=5$, both the HOMO-LUMO gap and the cohesive energy changes
monotonously with the radius. The existence of direct band gap in
hydrogenated silicon nanotubes is quite important for the
utilization of these nanosturctures in building nanoscale
optoelectronic devices.

\section{CONCLUSIONS}
A series of finite and infinite hydrogenated silicon nanotubes are
systematically studied by performing first-principles calculations.
Our results reveal that one-dimensional stable SiNTs
$Si_{m(3k+1)}H_{2m(k+1)}$ can be built by stacking $Si_{4m}H_{4m}$
cagelike clusters along the central axis of the cage. These tubes
have large cohesive energies and HOMO-LUMO gaps.  Among all the
tubes, those with the sizes of $m$=5 and 6 are the most stable,
because their $sp^{3}$ bond angle is most close to 109.5$^{\circ}$.
Thermodynamics analysis shows that these tubes may exist at room
temperature, which further confirms the stability of the proposed
silicon nanotubes. The infinite silicon nanotubes have also been
investigated and it is found that there is a direct-indirect-direct
band gap change with the increasing radius. For $m=4$, the direct
gap opens at Z point, while for $m\geq6$, the direct gap opens at
$\Gamma$ point. In the $m=5$ case, an indirect gap is observed.

We want to note that, although single wall silicon nanotubes have
attracted great attention recently, such silicon nanotubes have not
been synthesized yet experimentally due to the $sp^{3}$
hybridization of silicon and the subsequent unsaturated dangling
bonds. Our study indicates that hydrogen passivation may be a good
way to stabilize hollow single wall silicon nanotubes. Particularly,
necklace-like hollow structures built from single cage-like clusters
are proposed in our work. These findings may provide some useful
suggestions for experimental fabrications of hollow single wall
silicon nanotubes. \vspace{5mm}

\section*{ACKNOWLEDGMENTS}

This work was supported by the National Science Foundation of China
under Grants No. 10774148 and No. 10904148, the special Funds for
Major State Basic Research Project of China (973) under grant No.
2007CB925004. Some of the calculations were performed at the Center
for Computational Science of CASHIPS.

\begin{small}

\begin{figure}
\centering
\includegraphics[bb =40 0 900 790,width=1.60\textwidth]{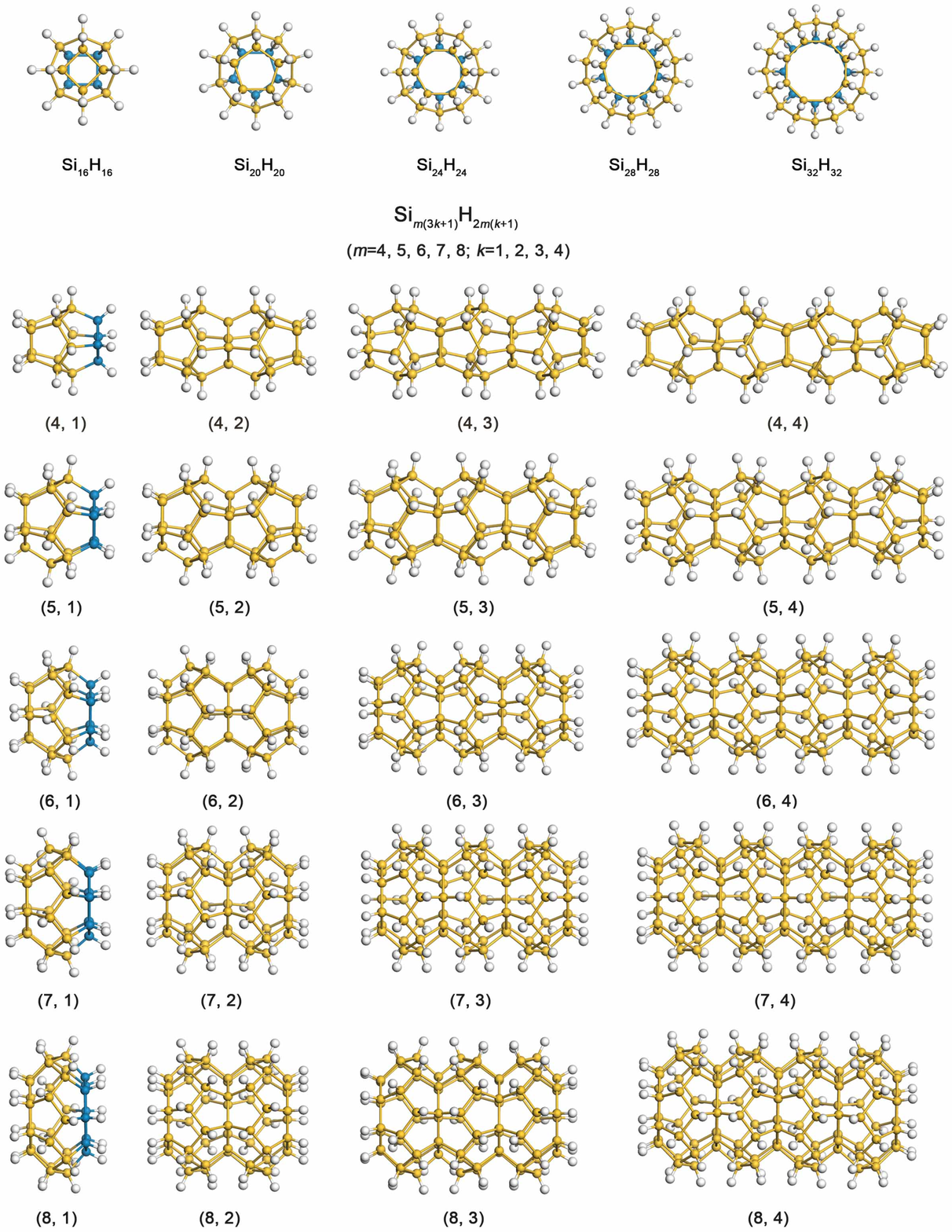}
\caption{\label{fig1}}
\end{figure}

\newpage
\begin{figure}
 \centering
\includegraphics[bb =0 0 750 800,width=1.20\textwidth]{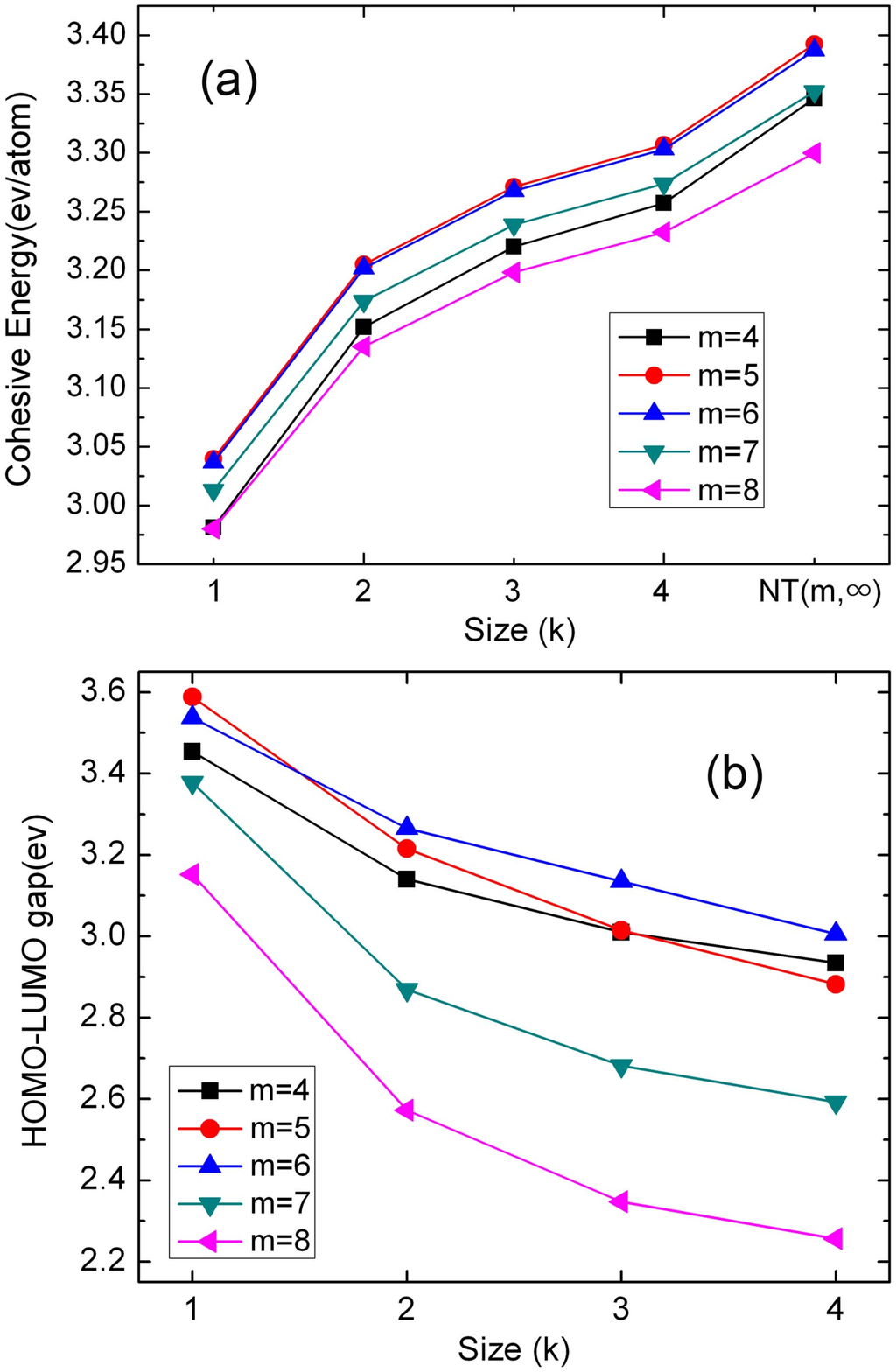}
\caption{\label{fig2}}
\end{figure}

\newpage
\begin{figure}
 \centering
\includegraphics[bb =40 0 620 650,width=1.30\textwidth]{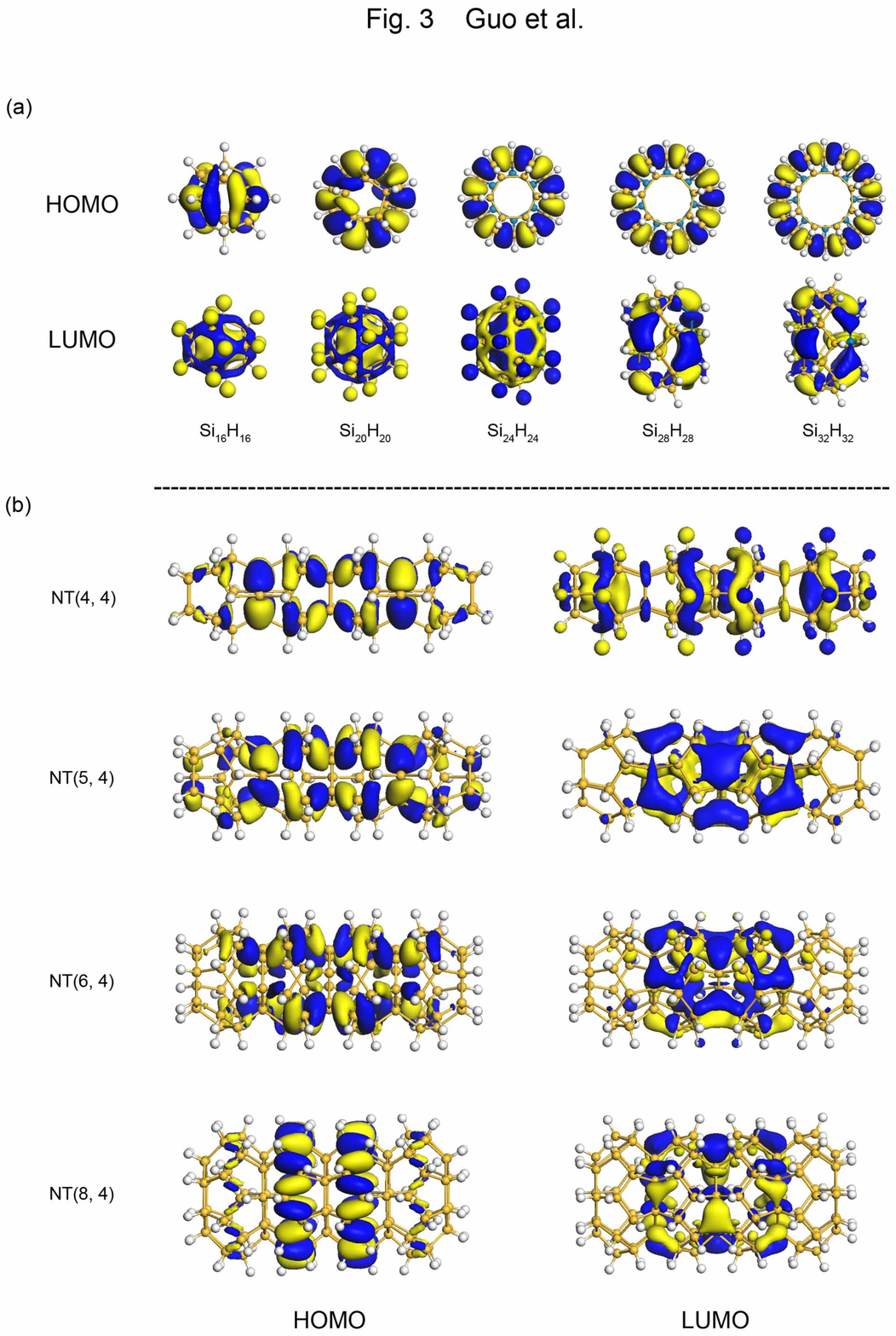}
\caption{\label{fig3}}
\end{figure}

\newpage
\begin{figure}
 \centering
\includegraphics[bb =50 0 700 500,width=1.50\textwidth]{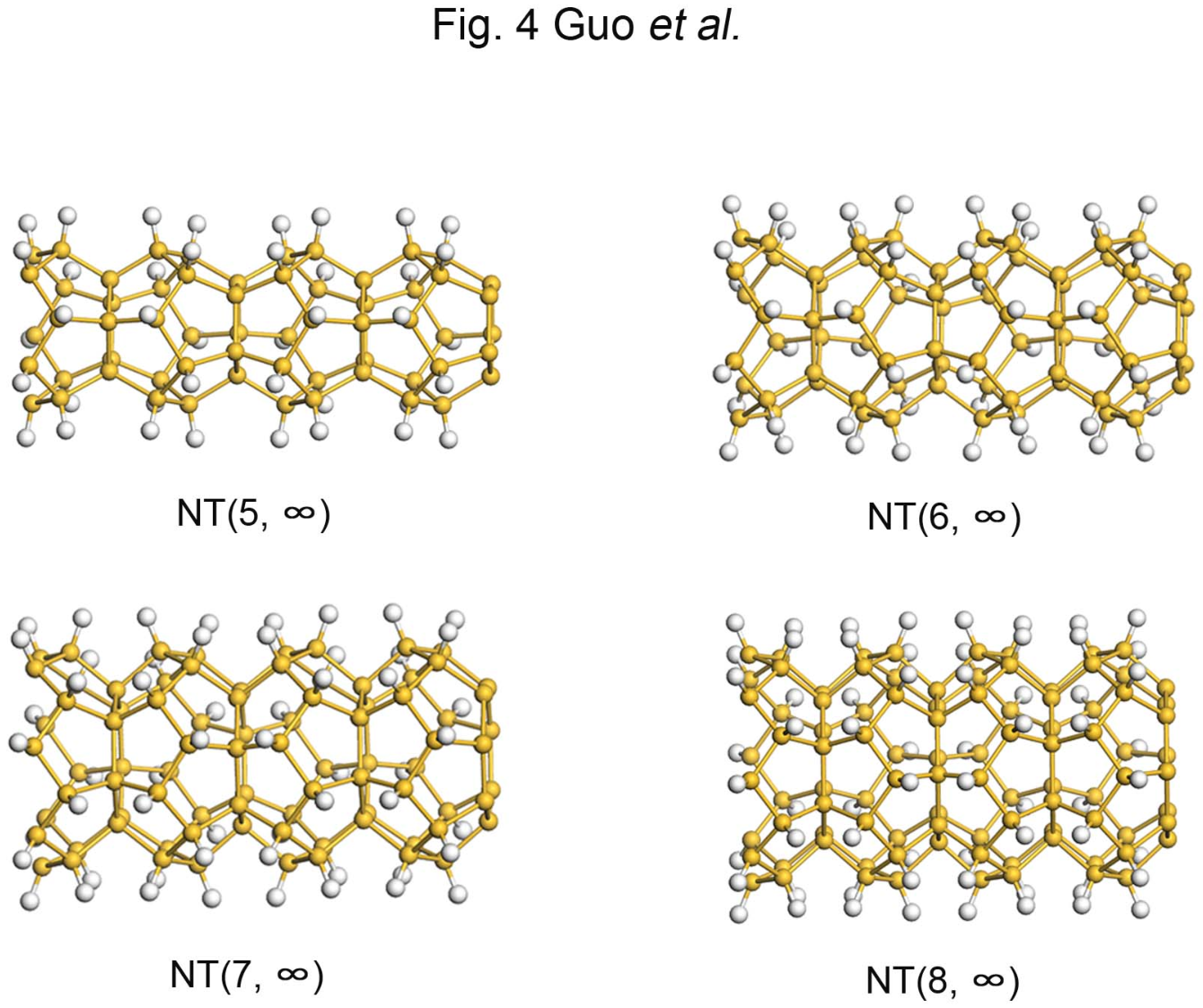}
\caption{\label{fig4}}
\end{figure}

\newpage
\begin{figure}
 \centering
\includegraphics[bb = 30 0 480 500,width=1.00\textwidth]{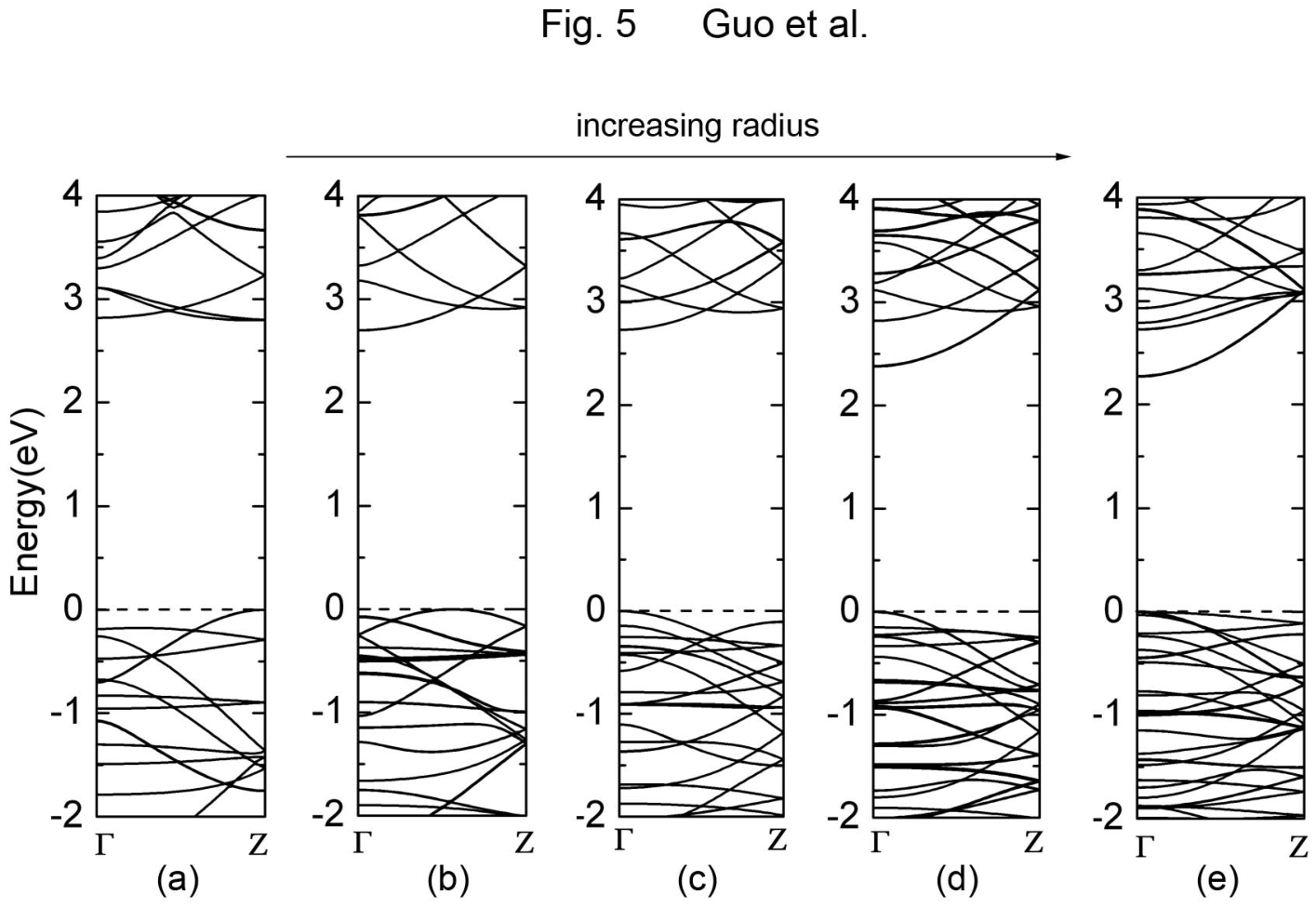}
\caption{\label{fig5}}
\end{figure}

\end{small}
\end{document}